\documentclass{emulateapj} 
\usepackage{apjfonts}              
\usepackage{graphicx}
\usepackage{amsmath}

\newcommand{\Msun}{\rm M_{\odot}}

\newcommand{\MBH}{M_{\mathrm{BH}}}
\newcommand{\MCMO}{M_{\mathrm{CMO}}}
\newcommand{\Mbulge}{M_{\mathrm{bulge}}}
\newcommand{\Mgal}{M_{\mathrm{gal}}}
\newcommand{\msigma}{\MBH\mathrm{-}\sigma}
\newcommand{\mcmosigma}{\MCMO\mathrm{-}\sigma}
\newcommand{\mbulge}{\MBH\mathrm{-}\Mbulge}
\newcommand{\mcmo}{\MCMO\mathrm{-}\Mgal}
\newcommand{\mcmogal}{\MCMO \approx 0.002\, \Mgal}
\newcommand{\re}{R_{\rm e}}

\slugcomment{Accepted to ApJ}

\begin{document}   
                       
\title{Correlations Between Central Massive Objects And Their Host
  Galaxies: \\
  From Bulgeless Spirals to Ellipticals} 

\author{Yuexing Li\altaffilmark{1}, Zolt\'an Haiman\altaffilmark{2},
  Mordecai-Mark Mac Low\altaffilmark{3}}  

\affil{$^{1}$Institute for Theory and Computation, Harvard-Smithsonian Center
  for Astrophysics, Harvard University, 60 Garden Street, Cambridge, MA 02138}
\affil{$^{2}$Department of Astronomy, Columbia University, New York,
NY 10027}
\affil{$^{3}$Department of Astrophysics, American Museum of Natural
History, 79th Street at Central Park West, New York, NY 10024}
\email{yxli@cfa.harvard.edu, zoltan@astro.columbia.edu, mordecai@amnh.org} 

\begin{abstract}
Recent observations reveal that a majority of galaxies contain a central
massive object (CMO), either a supermassive black hole (SMBH) or a compact
stellar nucleus, regardless of the galaxy mass or morphological type.
The masses of these CMOs correlate tightly with those of the host galaxies,
$\mcmogal$. Several recent studies argue that
feedback from black holes can successfully explain the $\msigma$ correlation
in massive elliptical galaxies that contain SMBHs. However, puzzles remain in
spirals or dwarf spheroids that do not appear to have black holes but instead
harbor a compact central stellar cluster. Here we use three-dimensional,
smoothed particle hydrodynamics simulations of isolated galaxies to study the
formation and evolution of CMOs in bulgeless disk galaxies, and simulations of
merging galaxies to study the transition of the CMO-host mass relation from
late-type bulgeless spirals to early-type ellipticals. In the simulations,
absorbing sink particles represent 
either SMBHs or star clusters, while stellar feedback on the gas is
implemented by assuming an isothermal equation of state with effective sound
speed of order 10~km~s$^{-1}$.  Our simulations show that the mass of the CMO
correlates with that of the host galaxy in both isolated bulgeless spirals and
in ellipticals formed through mergers, and that $\MCMO$ correlates with the
global star-formation efficiency in the galaxy. We find that the final mass of
the CMO is dominated by the accreted mass, rather than the initial fragment
mass.  The denser nuclei of more massive galaxies have higher mass accretion
rates, and both the final accreted CMO mass and the recently formed 
stellar mass increase monotonically with the total mass of the galaxy.
Our results suggest that the observed correlations may be established
primarily by the depletion of gas in the central region by accretion and  
star-formation, and may hold for all galaxy types.  A systematic
search for CMOs in the nuclei of bulgeless disk galaxies would offer a
test of this conclusion.
\end{abstract}


\section{INTRODUCTION}
Supermassive black holes (SMBHs) appear to exist in most, if not all,
galaxies (see, e.g., the reviews by \citealt{Haiman2004} and
\citealt{Ferrarese2005}). Over the past few years, SMBHs have been detected in
about 40 nearby galaxies using gas and stellar dynamical methods (e.g.,
\citealt{Kormendy1995, Richstone1998, Ho1999, Merritt2001}). These galaxies
include nearby ellipticals and lenticulars (e.g., \citealt{Marconi1997,
Macchetto1997}), as well as spirals with bulges such as the Milky Way
(\citealt{Genzel1997, Ghez2000, Ghez2003, Schodel2002}).  Most of these galaxies
are luminous ones with total masses larger than $10^{12}\, \Msun$. There
appear to be tight correlations between the masses of the SMBHs,
$M_{\rm BH}$, and the global properties of the spheroid components of
their hosts, such as their luminosities and masses (\citealt{Magorrian1998,
  Marconi2003}), light concentration (\citealt{Graham2001}), and velocity
dispersions (\citealt{Ferrarese2000, Gebhardt2000, Tremaine2002}).

Recently, low- and intermediate-luminosity galaxies, with $M_{\rm B} \gtrsim
-18$ mag, have been studied by \cite{Wehner2006} using the Hubble Space
Telescope (HST) Wide-Field/Planetary Camera Dwarf Elliptical Galaxy Snapshot
Survey of the Leo Group and Virgo and Fornax clusters \citep*{Lotz2004}, and
by \citet{Cote2006} and \citet{Ferrarese2006}, using HST Advanced Camera for
Surveys data for $\approx$100 early-type galaxies in the Virgo cluster
\citep{Cote2004}. These analyses show that a majority of these faint galaxies
contain a compact stellar cluster at their center.  The masses of these star
clusters correlate tightly with those of their host galaxies, $\mcmogal$, 
following the same relation between SMBH and galaxy masses seen in more
luminous galaxies with $M_{\rm B} \lesssim -20$ mag \citep{Magorrian1998,
  Ferrarese2000, Marconi2003}. A correlation between the mass of 
nuclear star clusters and galaxy properties holds not only in elliptical
galaxies, but also in spiral galaxies with bulges \citep{Balcells2003,
Rossa2006}.  Nuclear star-clusters with masses in the range $\sim 10^6-10^8\,
\Msun$ have also been found in a handful of bulgeless spiral galaxies
\citep{Walcher2005}. These findings strongly suggest that the SMBHs in massive  
galaxies and the compact stellar nuclei in less luminous galaxies may form as
a result of similar physical processes, linked to the formation of the galaxy
as a whole. 

Many models have been proposed for the observed SMBH-bulge
correlations, as reviewed by \cite{Robertson2006B}, notably including
self-regulation by global feedback, which suppresses further accretion and
star formation (e.g., \citealt{Silk1998, Haehnelt1998, Fabian1999, King2003,
  Wyithe2003, DiMatteo2005, Springel2005A, Sazonov2005, Murray2005,
  Wyithe2005}); as well as  environmental regulation (e.g., \citealt{Burkert2001, 
MacMillan2002, Adams2001, Balberg2002, Miralda2005, Begelman2005}); and star
formation models in which gas dissipation affects 
the growth of the SMBHs (e.g., \citealt{Archibald2002, DiMatteo2003,
Kazantzidis2005}). In particular, self-regulated models with SMBH
feedback of photoionization and Compton heating \citep{Sazonov2005}, or 
in the form of thermal energy coupled to the ambient gas have been
demonstrated to successfully reproduce many observed properties of elliptical
galaxies formed by major mergers, including the $\msigma$ relation
(\citealt{DiMatteo2005, Robertson2006A}), galaxy colors
\citep*{Springel2005B}, X-ray gas emission \citep{Cox2006A}, quasar properties 
and luminosity functions \citep{Hopkins2006A, Hopkins2006B}, as well as the
luminous quasars observed at the highest redshift \citep{Li2006A}.

However, self-regulated models are applicable primarily to galaxies with
SMBHs (i.e., massive elliptical galaxies). The new observations by
\cite{Cote2006}, \cite{Ferrarese2006} and \cite{Wehner2006} show that many
low- and intermediate-mass galaxies that do not appear to have SMBHs still 
have central compact stellar clusters that obey, to within observational
errors, an indistinguishable mass correlation with their host galaxy.  This
suggests that galactic black holes and central compact stellar nuclei,
hereafter collectively referred to as compact massive objects (CMOs), 
a terminology first introduced by \cite{Cote2006}, may have formed from the
same processes --- gravitational collapse of a massive clump of gas at the
center of the galaxy. Furthermore, it suggests that the CMO-host mass
correlation may be universal, regardless of galaxy type or mass. 

In order to test this hypothesis, we start our investigation of CMOs
with bulgeless spirals, which are the simplest galaxy models, and
therefore ideal to study the physical process of CMO
formation. Nuclear star clusters have been observed in bulgeless
spirals by \cite{Boker2002, Boker2004} and \cite{Walcher2005,
  Walcher2006}. These authors find that the growth of the CMOs is closely 
linked to the star formation history of the host galaxy. We here examine the
CMO-host mass correlation in simulations of isolated, bulgeless disk galaxies
with a wide range of masses and gas fractions. We then further investigate how
the CMO-host relation changes in simulations of mergers between spiral galaxies.

The rest of this paper is organized as follows. In \S~2, we describe
our computational methods, galaxy models and parameters. In \S~3, we
study the formation and evolution of CMOs in isolated, bulgeless,
spiral galaxies. We then study the formation of CMOs in mergers
between disk galaxies in \S~4. We discuss the connections between CMOs
and host galaxies in both spirals and ellipticals, and conclude in
\S~5.

\section{COMPUTATIONAL METHOD}
\label{sec_com}

We use the publicly available, three-dimensional, parallel,
N-body/smoothed particle hydrodynamics (SPH) code GADGET v1.1
\citep{Springel2001}, modified to include absorbing sink particles
\citep*{Bate1995} to directly measure the mass of gravitationally
collapsing gas. \citet*{Li2005A} and \citet{Jappsen2005} give detailed
descriptions of sink particle implementation and interpretation.  Sink
particles replace gravitationally bound regions of converging flow
that reach number densities $n_{\rm sink} > 10^3$~cm$^{-3}$. The sink
particles have a control volume with a fixed radius of 50 pc from which they
absorb surrounding bound gas.  They interact gravitationally and inherit the
mass, and linear and angular momentum of the gas they accrete.

\subsection{Galaxy Models}

\begin{deluxetable}{lcccccccc}
\label{tab1}
\tablecolumns{9}
\tablecaption{Spiral Galaxy Models and Numerical Parameters}
\tablehead{
\colhead{Model\tablenotemark{a}}       & 
\colhead{$R_{200}$\tablenotemark{b}}   &
\colhead{$M_{200}$\tablenotemark{c}}   & 
\colhead{$f_{\rm g}$\tablenotemark{d}} &
\colhead{$R_{\rm d}$\tablenotemark{e}} &
\colhead{$h_{\rm g}$\tablenotemark{f}} &
\colhead{$m_{\rm g}$\tablenotemark{g}} &
\colhead{LT\tablenotemark{h}} &
\colhead{HT\tablenotemark{i}}
}  
\startdata
SG50-1  &   71.43  &  4.15  &  0.2  &  1.41 & 10 &  0.21  & Y  &  N  \\
SG50-2  &   71.43  &  4.15  &  0.5  &  1.41 & 10 &  0.21  & Y  &  N  \\
SG50-3  &   71.43  &  4.15  &  0.9  &  1.41 & 10 &  0.37  & Y  &  N  \\
SG50-4  &   71.43  &  4.15  &  0.9  &  1.07 & 10 &  0.75  & Y  &  Y  \\
 & & & & & & &   &   \\   	     		   	   	    	    
SG100-1 &   142.86 &  33.22 &  0.2  &  2.81 & 10 &  0.66  & Y  &  N  \\
SG100-2 &   142.86 &  33.22 &  0.5  &  2.81 & 10 &  1.65  & Y  &  N  \\
SG100-3 &   142.86 &  33.22 &  0.9  &  2.81 & 10 &  2.97  & Y  &  Y  \\
SG100-4 &   142.86 &  33.22 &  0.9  &  2.14 & 20 &  5.94  & Y  &  Y  \\	
 & & & & & & &   &   \\	     	 			   	    	     
SG120-3 &   171.43 &  57.4  &  0.9  &  3.38 & 20 &  5.17  & N  &  Y  \\
SG120-4 &   171.43 &  57.4  &  0.9  &  2.57 & 30 &  10.3  & N  &  Y  \\
 & & & & & & &   &   \\	     	 			   	    	     
SG160-2 &   228.57 &  136.0 &  0.5  &  4.51 & 20 &  6.80  & N  &  Y  \\
SG160-3 &   228.57 &  136.0 &  0.9  &  4.51 & 30 &  12.2  & N  &  Y  \\ 
SG160-4 &   228.57 &  136.0 &  0.9  &  3.42 & 40 &  16.3  & N  &  Y  \\ 	
& & & & & & &   &   \\	     	                  	   	    
SG220-1 &   314.29 &  353.7 &  0.2  &  6.20 & 20 &  7.07  & Y  &  Y  \\
SG220-2 &   314.29 &  353.7 &  0.5  &  6.20 & 30 &  14.8  & N  &  Y  \\
SG220-3 &   314.29 &  353.7 &  0.9  &  6.20 & 40 &  15.9  & N  &  Y  \\
SG220-4 &   314.29 &  353.7 &  0.9  &  4.71 & 40 &  16.0  & N  &  Y  \\
\enddata 
\tablenotetext{a}{Model of single disk galaxy. First number is rotational
  velocity in km s$^{-1}$ at the virial radius, the second number indicates
  sub-model. Sub-models have varying fractions  $m_{\rm d}$ of total 
  halo mass in their disks, and given values of $f_{\rm g}$. Sub-models 1--3
  have $m_{\rm d}=0.05$, while sub-model 4 has $m_{\rm d}=0.1$. } 
\tablenotetext{b}{Virial radius in kpc within which the mean mass density of
  the halo is 200 times of the critical density.}
\tablenotetext{c}{Virial mass of the galaxy in 10$^{10}\, \Msun$.}
\tablenotetext{d}{Fraction of disk mass in gas.}
\tablenotetext{e}{Radial disk scale length in kpc where stellar
  surface density drops by $e^{-1}$.}
\tablenotetext{f}{Gravitational softening length of gas in pc.}
\tablenotetext{g}{Gas particle mass in units of $10^4\, \Msun$.} 
\tablenotetext{h}{Model run with LT mode  $c_s = 6$~km~s$^{-1}$.} 
\tablenotetext{i}{Model run with HT mode  $c_s = 15$~km~s$^{-1}$.}
\end{deluxetable}

\begin{deluxetable}{lcccccccc}
\label{tab1}
\tablecolumns{9}
\tablecaption{Galaxy Mergers and Numerical Parameters}
\tablehead{
\colhead{Model\tablenotemark{a}}       & 
\colhead{$V_{200} (1)$\tablenotemark{b}}   &
\colhead{$M_{200} (1)$\tablenotemark{c}}   &
\colhead{$V_{200} (2)$\tablenotemark{d}}   &
\colhead{$M_{200} (2)$\tablenotemark{e}}   & 
\colhead{$f_{\rm g}$\tablenotemark{f}} &
\colhead{$h{\rm g}$\tablenotemark{g}}  &
\colhead{$m_{\rm g}$\tablenotemark{h}} &
\colhead{$c_s$\tablenotemark{i}}} 
\startdata	   	  		    
MG1 &  50  & 4.15  & 50  & 4.15  & 0.2  & 10 &  0.37  & 6  \\
MG2 &  100 & 33.22 & 100 & 33.22 & 0.2  & 10 &  0.66  & 6  \\
MG3 &  100 & 33.22 & 100 & 33.22 & 0.9  & 10 &  2.97  & 6  \\
MG4 &  160 & 136.0 & 202 & 272.0 & 0.1  & 30 &  16.0  & 10 \\
MG5 &  220 & 353.7 & 220 & 353.7 & 0.2  & 40 &  20.0  & 6  \\
\enddata 
\tablenotetext{a}{Model of galaxy mergers. Note all progenitors have disk mass
  fraction of $m_{\rm d}=0.05$, . } 
\tablenotetext{b}{Rotational velocity in km s$^{-1}$ at the virial radius for
  the first progenitor.} 
\tablenotetext{c}{Virial mass of the first progenitor in 10$^{10}\, \Msun$.}
\tablenotetext{d}{Rotational velocity in km s$^{-1}$ at the virial radius for
  the second progenitor.}
\tablenotetext{e}{Virial mass of the second progenitor in 10$^{10}\, \Msun$.}
\tablenotetext{f}{Gas fraction of the progenitors.}
\tablenotetext{g}{Gravitational softening length of gas in pc.}
\tablenotetext{h}{Gas particle mass in units of $10^4\, \Msun$.} 
\tablenotetext{i}{Sound speed for the gas in km~s$^{-1}$.} 
\end{deluxetable}

Our models of isolated, bulgeless spiral galaxies consist of a dark matter
halo following the prescription of \citep*{Navarro1997}, and an initially
exponential disk of stars and isothermal gas.  The galaxy structure is based
on the analytical work by \citet*{Mo1998}, as implemented numerically by
\citet{Springel1999} and \citet{Springel2000}. The detailed description of the
models and numerical simulations are given in \cite{Li2005A}. Table~1 lists
the most important model properties and numerical parameters. In order to
sufficiently resolve gravitational collapse, the models are set up to
satisfy three numerical criteria: the Jeans resolution criterion
(\citealt{Bate1997, Whitworth1998}), the gravity-hydro balance criterion
for gravitational softening (\citealt{Bate1997}), and the equipartition
criterion for particle masses \citep{Steinmetz1997}. We choose the
particle number for each model such that they not only satisfy the
criteria, but also all runs have at least $10^6$ total particles. The
gas, halo and disk particles are distributed with number ratio $N_{\rm
g} : N_{\rm h} : N_{\rm d}$ = $5 : 3 : 2$. The gravitational softening
lengths of the halo $h_{\rm h} = 0.4$~kpc and disk $h_{\rm d}=
0.1$~kpc, while that of the gas $h_{\rm g}$ varies with models. The
minimum spatial and mass resolutions in the gas are given by $h_{\rm
g}$ and twice the kernel mass ($\sim 80 m_g$). We adopt typical values
for the halo concentration parameter $c = 5$, spin parameter $\lambda
= 0.05$, and Hubble constant $H_0 = 70$ km s$^{-1}$ Mpc$^{-1}$
\citep{Springel2000}.  

To test the effects of feedback, we have chosen two values of the isothermal
sound speed from the range of velocity dispersions typically observed for
neutral gas in disk galaxies 6--15~km~s$^{-1}$ \citep{Elmegreen2004,
Dib2006}. Models designated as low-temperature (LT) have $c_s =
6$~km~s$^{-1}$, while high-temperature (HT) models have $c_s = 15$ km
s$^{-1}$. We should emphasize that the effective temperature approach we use
in our simulations should be regarded as a simple approximation of the
underlying processes maintaining the interstellar turbulence level, including
stellar feedback, magnetorotational instability, and other processes
(see \citealt{mk04} for a review). All of the galaxy models have been
simulated in the HT mode. However, in our simulations, low mass HT models 
(e.g., some of the SG50 and SG100 models) do not appear to collapse in the
first 3 Gyrs. Only a limited number of runs have been performed in LT mode, as
it is prohibitively expensive to run high-mass LT models because they require
$\sim 10^7$ particles to fully satisfy the three numerical criteria. The model
properties and numerical parameters are summarized in Table~1.

The same set of simulations of isolated galaxies has successfully
reproduced many observations of star formation in disk galaxies,
including distributions and morphologies \citep{Li2005A}, and both global
and local Schmidt laws \citep*{Li2006B}, demonstrating that star formation in
galaxies are controlled by gravitational instability \citep*{Li2005B}. In
this paper, we focus on the connection between the CMO and the host
galaxy.

In order to investigate CMOs in elliptical galaxies, we have also done
five representative models of mergers of spiral galaxies that result in
elliptical galaxies. These mergers vary in the mass ratio of the merging
pair, progenitor size, gas fraction, effective sound speed, bulge component,
and orbital parameters and configuration. Model MG4 resembles the 
merger of Milky Way and Andromeda, in which the progenitor properties and
orbital parameters are set up following \cite*{Dubinski1996} and
\cite{Springel2000}. The mass ratio of the progenitors is roughly 1:2, and
each progenitor has a bulge with a mass that is 25\% of the disk mass, but neither
progenitor initially has a CMO.  The two galaxies are on a parabolic orbit,
with an initial separation of 700 kpc, and a pericentric distance of 5
kpc. The rest of the models are equal mass, head-on mergers of identical
bulgeless progenitors using models SG50-1, SG100-1, SG100-3, and SG220-1,
respectively. These mergers are run with parabolic orbits having pericentric
distance of 1~kpc, whose separation is the sum of the virial radii of the two
progenitors. Each of the mergers is run with particle number around $N_{\rm
  tot} \approx 2\times 10^6$. The model properties and numerical parameters of
the merger runs are listed in Table~2. We have also performed a resolution
study for the LT versions of models SG100-1 and SG220-1, running each with
total particle numbers of $10^5$, $8\times 10^5$, and $6.4 \times 10^6$, in
addition to the standard value of $10^6$.  These models cover a factor of four
in linear resolution. 

\subsection{Interpretation of Sink Particles as Clusters and Central Massive
  Objects} 

The absorbing sink particles in our models trace regions with pressure
of $P/k \sim 10^7$~K~cm$^{-3}$, and reach masses of $\gtrsim 10^6 \, \Msun$,
typical of star-forming proto-stellar clusters \citep{Elmegreen1997}.
We therefore interpret the formation of sink particles in general as
representing the formation of massive stellar clusters. We find in the
simulations, however --- both in the case of isolated galaxies and for
galaxy mergers --- that the most massive sink particle always resides
in the center of the isolated disk \citep{Li2005A} or the merger
remnant \citep*{Li2004}. It either forms there, or migrates to the
center quickly after formation due to dynamical friction. We therefore
distinguish this sink particle from the others,  and identify it 
as a CMO.

To quantify star formation, we assume that individual sinks represent dense
molecular clouds that form stars at some efficiency. Observations by
\citet{Rownd1999} suggest that the {\em local} star formation efficiency (SFE)
in molecular clouds remains roughly constant. \citet{Kennicutt1998} found SFE
of 30\% for starburst galaxies that \citet{Wong2002} showed are dominated by
molecular gas. We therefore adopt a fixed local SFE of 30\% to convert the
mass of sinks to stars, while making the simple approximation that the
remaining 70\% of the sink particle mass remains in gas form. This conversion
rate was shown to reproduce both the global and local Schmidt laws observed in
nearby galaxies \citep{Li2006B}.  

We assume, on the other hand, that the CMO inherits the total mass of
the central sink particle, for the following reasons. (1) If the CMO
forms a compact stellar cluster, then the star formation efficiency
could be $> 50\%$ due to extremely high pressure and density at the
galactic center \citep{Elmegreen1997}, where the deep gravitational
potential also helps to keep the gas bound the CMO. (2) If a fraction
of the CMO collapses into a black hole, then it could rapidly accrete
the remaining gas. (3) Although feedback-driven winds from either
supernovae or a quasar could expel the gas from galactic nuclei, the
effect may be minor. A study of galactic winds by \cite{Cox2006B}
shows that the mass loss by a typical stellar wind (wind
efficiency $\sim 0.5$, wind velocity $\sim 500$ km s$^{-1}$) is less
than 10\% of the initial total gas mass. The quasar outflow is
comparable to this in galaxies with accreting black
holes. Furthermore, most of the gas in the central region is consumed
quickly by either star formation or black hole accretion, so the mass
loss from the CMO by feedback is likely negligible \citep{Cox2006A}.

The final mass and fate of the CMO should depend on the galaxy
potential and dynamical evolution. The larger the galaxy, the more
massive its central clump, due to stronger gravitational instability
\citep{Li2005B}. The gas clump may then form a supermassive black hole
directly by rapid core collapse within a deep potential well, fragment
to form a massive star cluster in a less-massive galaxy, or engage in
cluster-cluster collisions to form a SMBH in galaxy mergers due to
stellar dynamical processes (see, e.g., \citealt{Rees1984} and
\citealt{Haiman2004} for reviews of different routes to forming a
SMBH). Furthermore, the $\mcmo$ relation may be established by the
same physical mechanism that determines the fate of the clump ---
massive galaxies or galaxy mergers that have higher CMO mass tend to
form stars at a higher rate than less massive ones. These
considerations suggest that the $\mcmo$ relation is directly connected
to the star formation efficiency of the galaxy.

In the simulations, we do not resolve the transition within the sink
particle from a collapsed gas clump to a black hole or a star
cluster, and consequently we will limit our discussion, and
comparisons to observations, to CMOs in general.

\section{Central Massive Objects in Bulgeless Spiral Galaxies}

\subsection{The Formation and Evolution of CMOs}

\begin{figure}
\begin{center}
\includegraphics[width=3.6in]{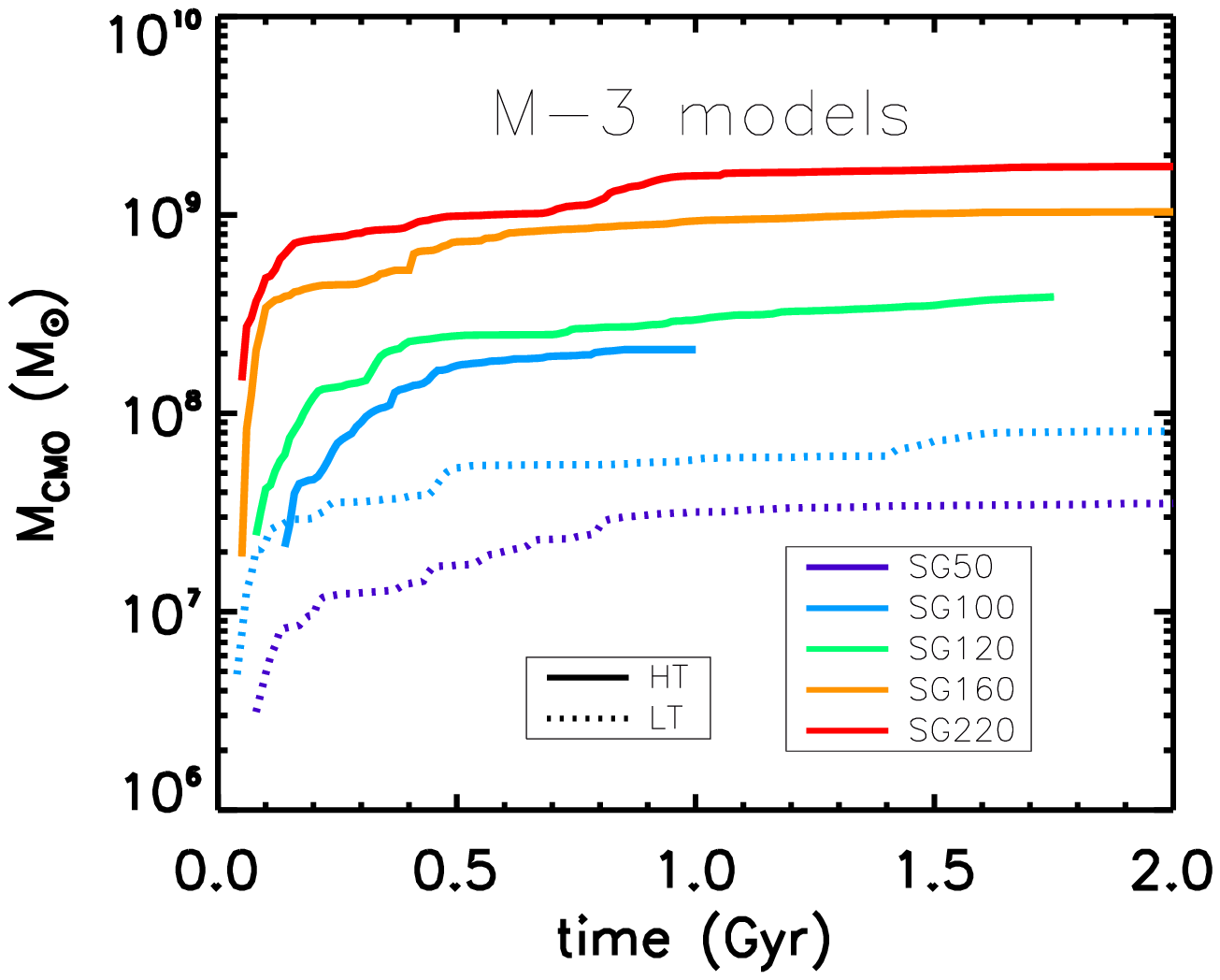}
\includegraphics[width=3.6in]{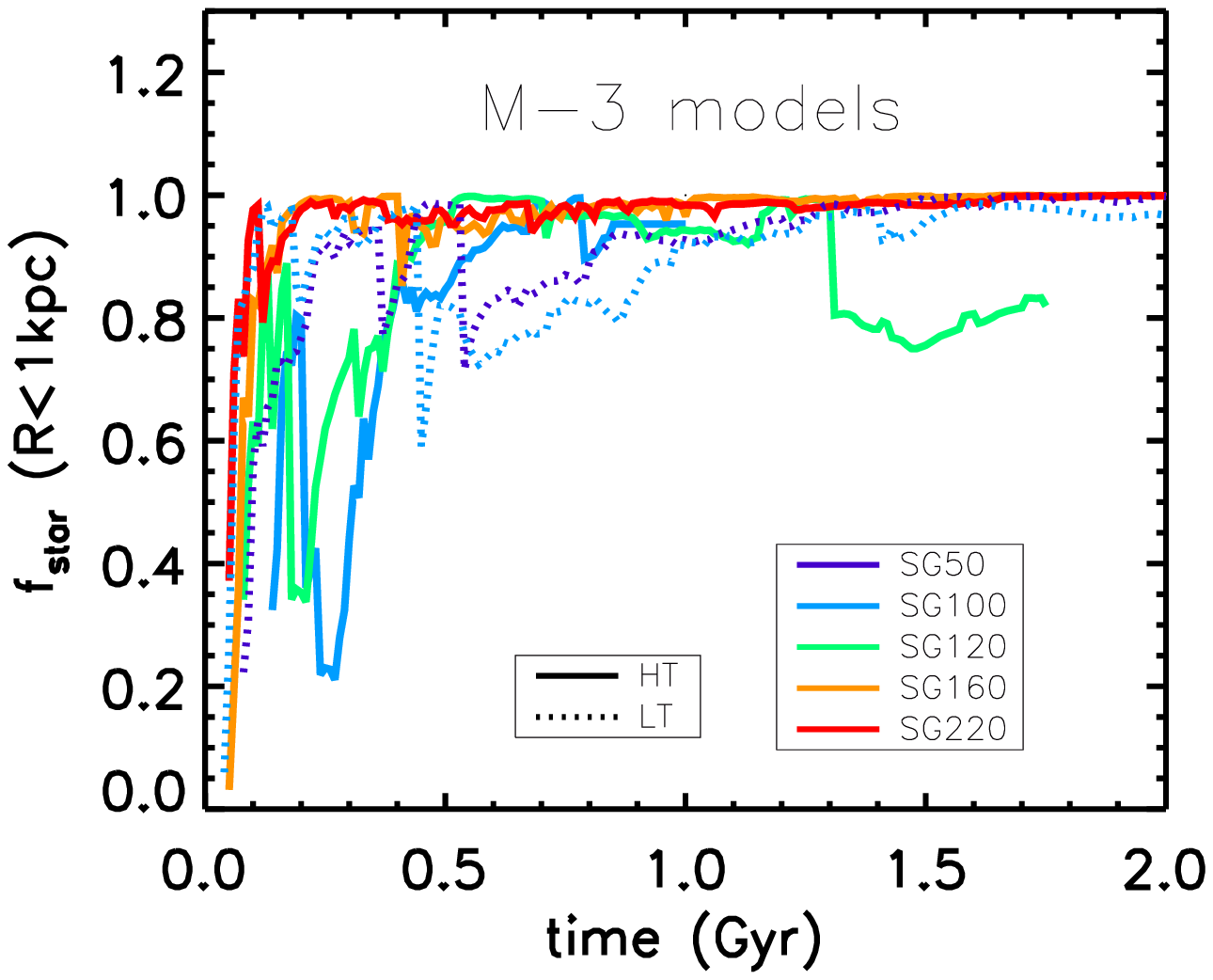}
\includegraphics[width=3.6in]{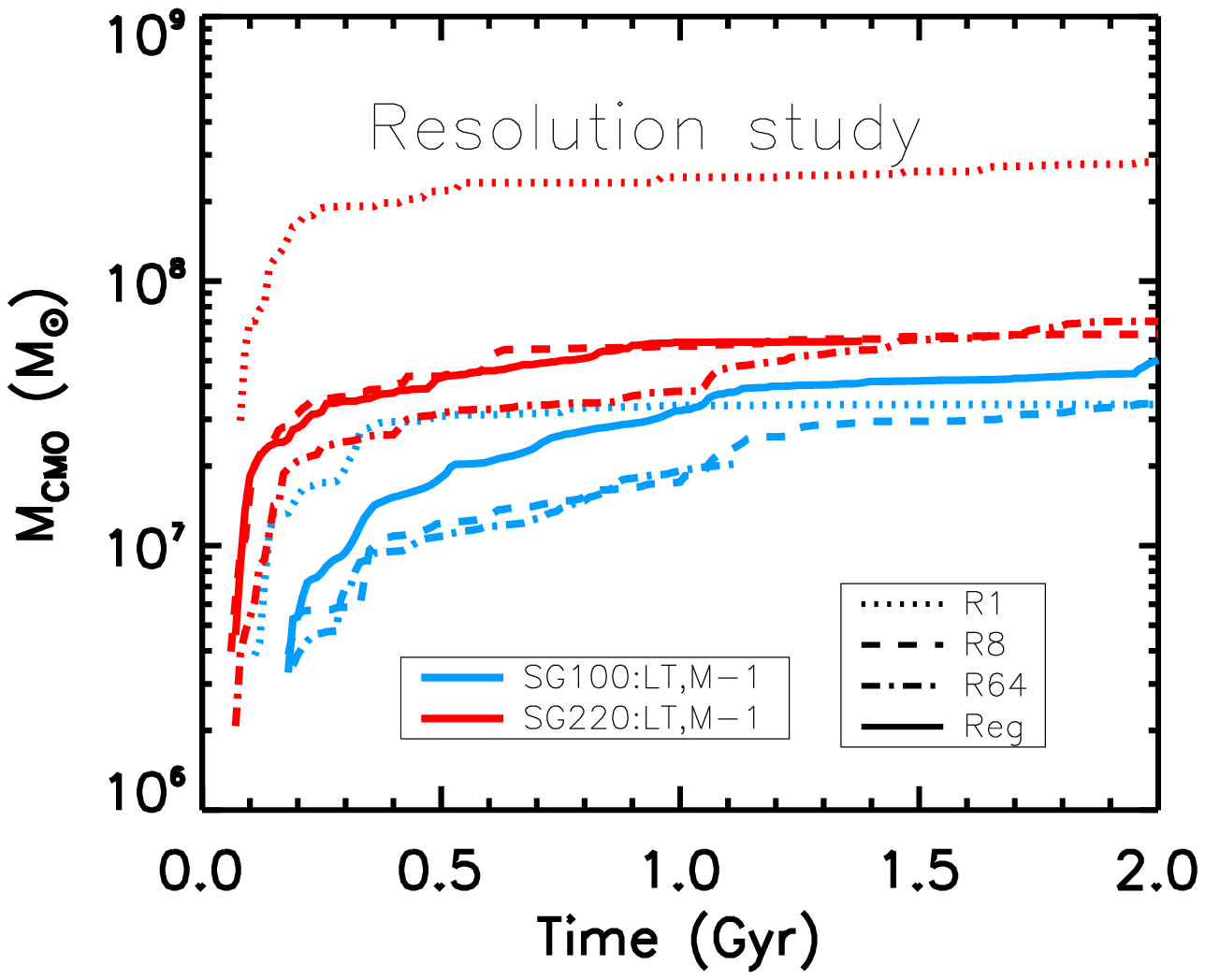}
\vspace{0.1cm}
\caption{\label{fig_mcmo_t} Mass growth histories of the CMOs in selected
  galaxy models (top panel), and the stellar mass fraction in the
  nuclear region within 1 kpc of the galactic center (middle
  panel). Also shown are the histories of CMO mass from  the resolution
  study (bottom panel). The designations R1, R8, R64 and Reg indicate
  total particle numbers of $10^5$, $8\times 10^5$, $6.4\times10^6$,
  and $10^6$, respectively.}
\end{center}
\end{figure}

\begin{figure*}
\begin{center}
\includegraphics[width=7.2in]{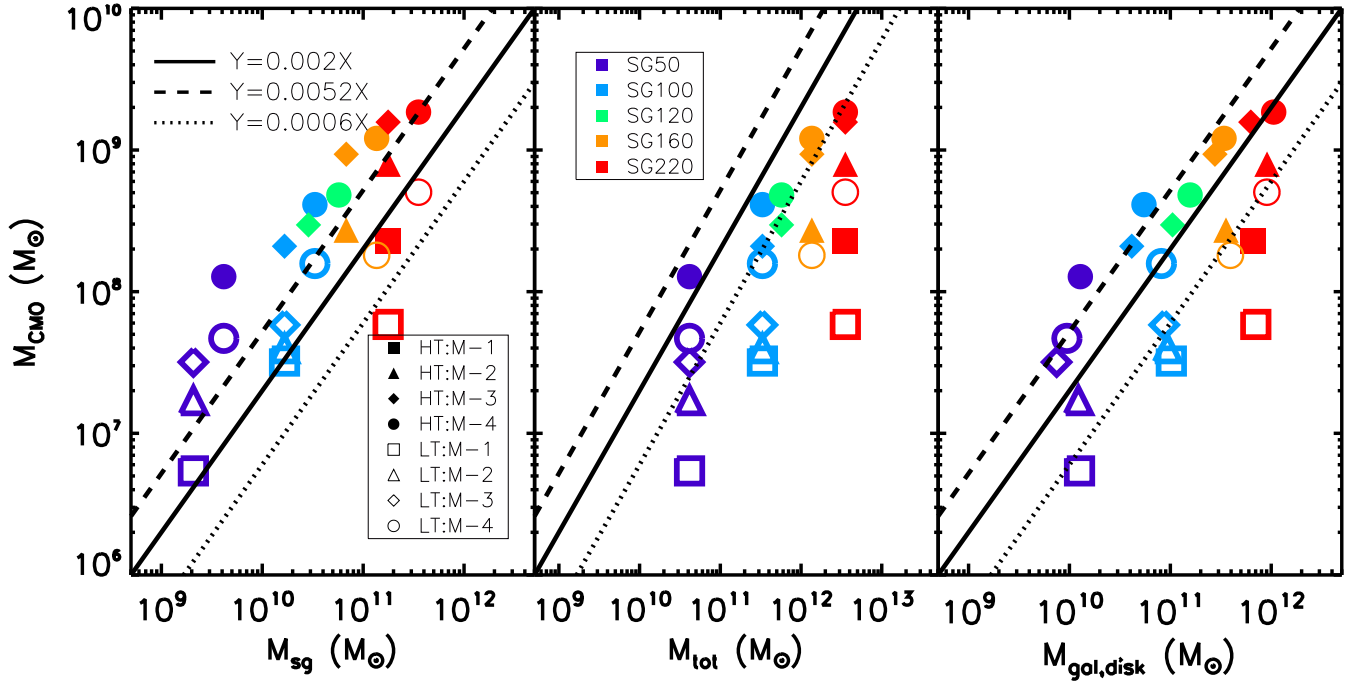}
\vspace{1cm}
\caption{ The relation between CMO mass and host mass in simulated bulgeless
spiral galaxies. Galaxies are shown from both the LT ({\em open symbols}) and
HT ({\em filled symbols}) models. The color and the shape of the symbol
indicates the rotational velocity and sub-model (gas fraction and disk scale
length) for each galaxy described in Table~1. The solid line is the
best fit to spheroid systems observed by \cite{Ferrarese2006}, while the
dotted line and the dashed line indicate the observed range. We show three
different characterizations of the host mass, as discussed in the text: ({\em
left}) $M_{\rm sg}$ the mass of both stars and gas; ({\em middle}) $M_{\rm
  tot}$ the total mass; and ({\em right}) $M_{\rm gal, disk}$ the
total mass within the half-mass radius. The last definition produces a scaling
relation close to that observed between SMBHs and stellar spheroids
\citep{Ferrarese2006}, suggesting that it may be the closest analog of the
mass $M_{\rm gal}$ used by \cite{Cappellari2006} to characterize bulgeless
spirals. Note that the two LT models in {\em thin} circles, SG160-4 and
SD220-4 do not fully satisfy the Jeans criterion, but judging from our
resolution study, likely  represent upper limits to the CMO mass. See  
\S~\ref{cmo_spiral} for details. }
\label{fig_mcmo_mgal_sp}
\end{center} 
\end{figure*}

In all of our simulations, high gas density regions collapse to form
stars and clusters once the disk becomes gravitationally
unstable. Sink particles first form in the central regions of
galaxies; star formation subsequently propagates outwards. The
collapsed objects continue to grow by gas accretion.

The growth histories of CMOs in several models were shown in
Figure~\ref{fig_mcmo_t}. The CMO initially collapses at a mass above
the Jeans mass ($M_{\rm J} \sim 10^6 \, \Msun$ and $\sim 2\times 10^7
\, \Msun$ for LT and HT models, respectively). We find that in the first few
hundred million years, the CMO grows rapidly, accreting gas particles
gravitationally bound to it, at a rate that is within an order of magnitude of
the spherical Bondi-Hoyle accretion rate \citep{Hoyle1941, BondiHoyle1944,
  Bondi1952}: 
\begin{equation}
\label{eq_bondi}
\dot{M}_{\rm {BH}}  
   =  \frac{4\pi \alpha  G^2 M_{\rm CMO}^2 \rho}{(c_s^2 + v^2)^{3/2}}
\end{equation}
where $M_{\rm {CMO}}$ is the CMO mass, $\rho$ and $c_s$ are the 
density and sound speed of the gas, respectively, $\alpha$ is a dimensionless
parameter of order unity, and $v$ is the velocity of the CMO relative
to the gas. After the initial accretion period, the growth slows down. In
about 1 Gyr, the masses of most CMOs saturate due to the depletion of gas
around the nucleus by star-formation, as shown in the middle panel of
Figure~\ref{fig_mcmo_t}. We note that a study by  
\cite{Sazonov2005} of radiative feedback from accreting black holes in 
ellipticals shows that the observed $\msigma$ relationship could be
established once most ($> 99\%$) of the gas has been converted into
stars. This finding agrees with our results that the growth of the 
CMO is tied to the star formation in the galaxy.  

The resolution studies show that the masses of the CMOs converge to better
than a factor of two in runs with particle numbers $\ge 10^6$ for both models
SG100-1 and SG220-1, in which just the gas particle mass is smaller than the
Jeans mass at which the CMO collapses initially. We include in our analysis
two models that satisfy this requirement in order to extend the range of LT
models that we can consider, as we will discuss.

\subsection{The $\mcmo$ Correlation}
\label{cmo_spiral}

\begin{figure}
\begin{center}
\includegraphics[width=3.5in]{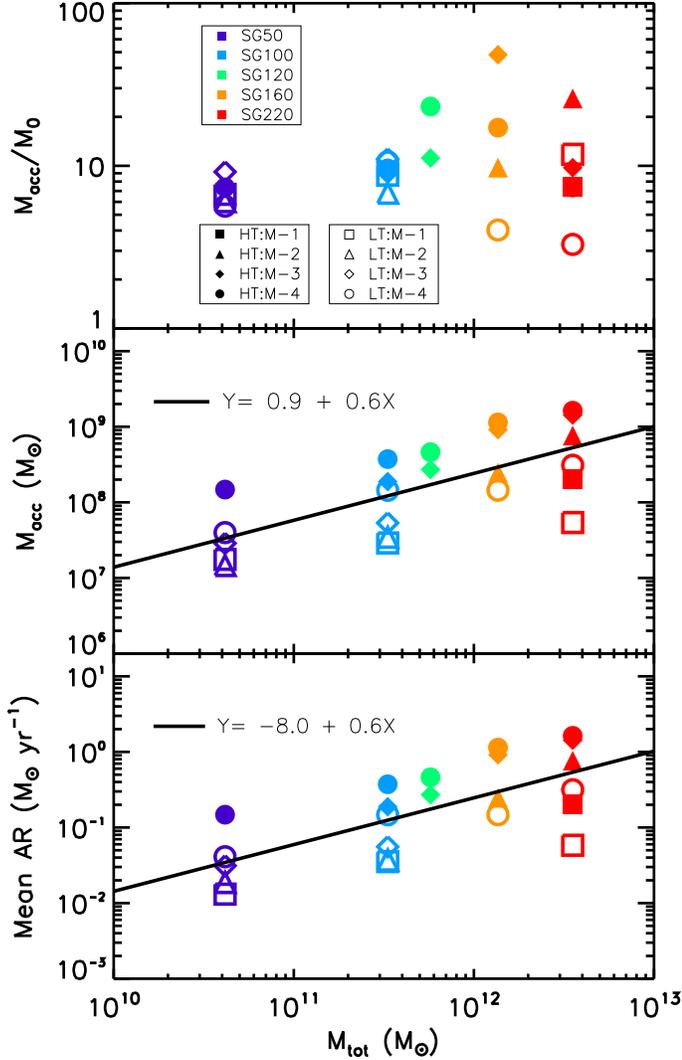}
\vspace{1cm}
\caption{\label{fig_macc_mgal} This figure shows how the CMOs gain
  mass, as a function of total galaxy mass $M_{\rm tot}$. The {\em top
  panel} shows the ratio of the accreted mass $M_{\rm acc}$ to the
  initial collapsed seed mass $M_0$, while {\em middle} and {\em
  bottom} panels show $M_{\rm acc}$ and mean accretion rate of the CMO,
  respectively, as functions of $M_{\rm tot}$ of the galaxy. The mean
  accretion rate is the average over 1 Gyr (or the maximum simulation time if
  shorter) when $\MCMO$ is measured, as explained in the text.  The symbols
  are the same as in the previous plots, while the solid line is the
  least-square fit to the data.} 
\end{center}
\end{figure}

In Figure~\ref{fig_mcmo_mgal_sp}, we show the correlation between the
mass of CMO and that of the host for the bulgeless, isolated disk
models. We consider three different definitions of galaxy
mass: $M_{\rm sg}$, the combined mass of stars and gas, 
$M_{\rm tot}$, the total mass of the galaxy including stars, gas,
and dark matter, and $M_{\rm gal, disk}$, the total
mass of the galaxy within 5$\re$, where $\re$ is the half-light
radius, containing half of the gas and stellar mass of the disk
component. In all three cases, $\MCMO$ is measured after 1 Gyr, or in
the few cases where the simulations stopped before 1 Gyr, at the last
time step. The two LT models SG160-4 and SD220-4 (shown in thin circles) do
not fully satisfy the Jeans criterion (i.e., the SPH kernel mass of 32
gas particles is smaller than the Jeans mass) due to computation
limit. However, from the resolution study in Figure~\ref{fig_mcmo_t},
the CMO mass converges to roughly a factor of two even
in runs where the gas particle mass is smaller than the initial CMO mass.  
We believe that these models represent the upper limits of the CMO mass of
these two models. 

Figure~\ref{fig_mcmo_mgal_sp} shows that for any definition of the
galaxy mass, the mass of CMO correlates strongly with the mass of its
host, despite the lack in our models of any explicit feedback from the
CMO.  It also appears that the correlation has a similar normalization
and linear slope to that found for both star clusters and SMBHs. 

In reality, the neutral gas velocity dispersion in individual galaxies 
typically has a range of $6 - 15$ km s$^{-1}$ so the scatter in the $\mcmo$
relation, represented by the combination of the LT and HT runs, may reflect the
range of effective temperature we use here. Because the masses of the CMOs
depend on the effective sound speed of the gas, it is desirable to have a wide
range of models with different effective sound speeds. Nevertheless,
from the models we have simulated, there is reasonable overlap of the two (LT
and HT) temperature regimes, as shown in 
Figure~\ref{fig_mcmo_mgal_sp}. This suggests that the $\mcmo$ relation is 
similar for different temperatures, apart from a factor of $\approx 2$ shift
in the overall normalization, independent of host mass.  

It would be interesting to compare these findings directly with the
observations, but such a comparison is difficult at
present. \cite{Walcher2005} have determined the masses of nuclear star
clusters in 9 bulgeless spiral galaxies, and found them to range between
$8\times 10^5 \, \Msun$ and $6\times 10^7 \, \Msun$.  These appear a 
factor of several smaller than the masses of the CMOs we find in
galaxies with $M_{\rm tot}=10^{11}-10^{12} \, \Msun$, but the
masses of the hosts in their sample are not discussed, and
their sample may also suffer from selection biases.

Other comparisons are possible, but have to be interpreted with
caution, since it is not clear how to relate the bulgeless disks in
these models to the spheroid components that are found to correlate
with CMOs in other observations.  Nevertheless, we present here two
such brief comparisons:

(1) \cite{Balcells2003} and \cite{Rossa2006} study late-type spirals with
bulges. They find that the masses of the CMO at fixed bulge luminosity is
about 3 times above the mass expected from the ${\MBH}$ vs. $L_{\rm bulge}$
relation \citep{Marconi2003}.  In the left panel of
Figure~\ref{fig_mcmo_mgal_sp}, we find a similar offset (relative to the
$\mbulge$ relation), if the total baryon mass is considered as a proxy for the
bulge mass.  One possible rationale for this comparison is that most of the
baryon content of our simulated bulgeless disk galaxy may represent the stars
making up the bulge of any galaxy that subsequently forms out of this
bulgeless spiral.  

(2) \cite{Cote2006, Ferrarese2006} and \cite{Wehner2006} study early-type
galaxies, and find a correlation between the masses of the CMOs and the host
galaxies.  In particular, Ferrarese et al. define $\Mgal$, following the
formula given by \cite{Cappellari2006}, as $\Mgal = \alpha \, \re \,
\sigma^2\, /G$, where $G$ is the gravitational constant, $\sigma$ is the
observed velocity dispersion, and $\alpha = 5$ is a constant.  This typically
represents $\sim 0.25$ of the total galaxy mass (see below).  Our
definition of $M_{\rm gal, disk}$ is designed to mimic this quantity,
and, as the right panel of Figure~\ref{fig_mcmo_mgal_sp} shows, we
find, within errors, that our bulgeless disk galaxies obey the same
correlation.

Both of the above comparisons suggest that the $\mcmo$ relation
established in our simulated bulgeless disk galaxies may evolve into
the observed correlation between $\MCMO$ and their spheroid host, once
the bulgeless disk galaxy (or a fraction of its mass) evolves into a
spheroid.  Establishing a direct connection, however, is not possible
without modeling in detail the further evolution of both the host
galaxy and the CMO (the latter, of course, may also gain mass).  We
next turn to CMOs that form in our simulations of merging galaxies.

\subsection{Toward a Physical Explanation of the $\mcmo$ Correlation}
\label{sec_physics}

\begin{figure}
\begin{center}
\includegraphics[width=3.5in]{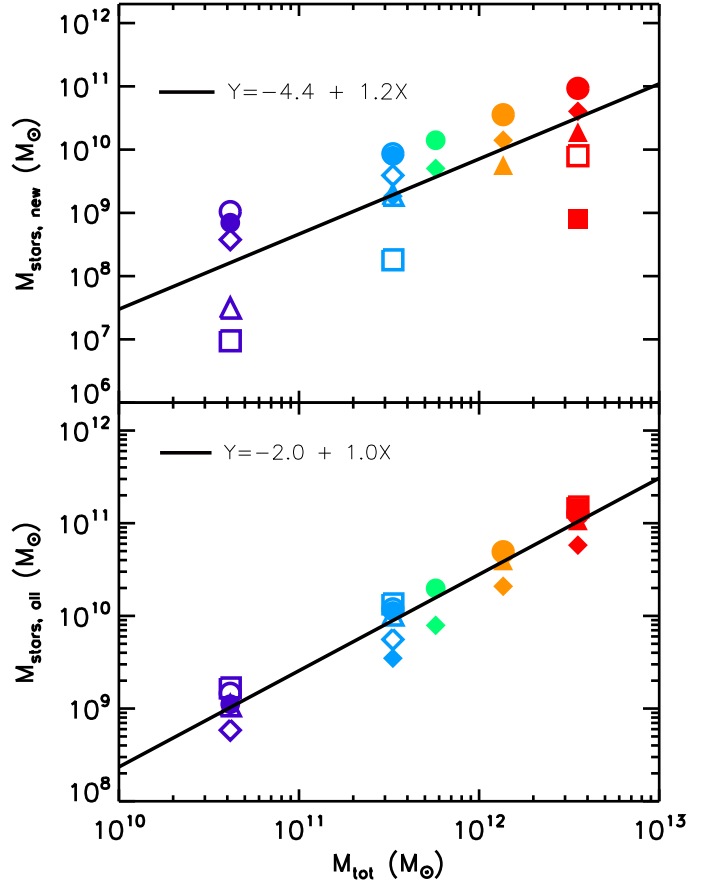}
\vspace{1cm}
\caption{\label{fig_mstar_mgal} Stellar mass of the galaxy as a function of total
  galaxy mass $M_{\rm tot}$. The {\em top panel} shows the mass of newly formed
  stars (30\% of the sink particles formed in the simulation, aside from the CMO), 
  while the {\em bottom  panel} shows the mass of all stars, including both
  newly formed ones and the old disk stars.}
\end{center}
\end{figure}

\begin{figure}
\begin{center}
\includegraphics[width=3.5in]{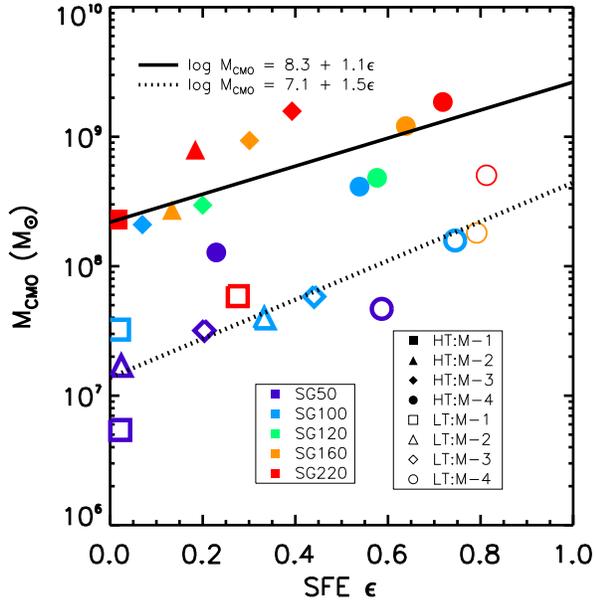}
\caption{\label{fig_mcmo_e} Correlation between the
  mass of the CMO and the global star formation efficiency in the
  galaxy in different models. Symbols are the same as in
  Figure~\ref{fig_mcmo_mgal_sp}. The black lines show a least-square fit to
  the data points from models of each temperature.  Because the CMO mass
  depends on the effective sound speed, so the lower Jeans masses in
  the LT runs result in a higher global SFE, and a smaller CMO. Our choice of
  two discrete values of $c_s$ represents the range of scattering (see text
  for more discussions). }  
\end{center}
\end{figure}

The physics responsible for the mass correlation, in general, must be
tied to the formation of a seed CMO, and its subsequent growth by
accretion.  To investigate this issue further, we define the initial
collapsed seed CMO mass as $M_0$, and the accreted mass as $M_{\rm
acc}$.  The final CMO mass is the sum of these two, $\MCMO=M_0+M_{\rm
acc}$, where $M_0$ is essentially the local Jeans mass. Our sink particles
form at a fixed,  arbitrary density $n_{\rm sink}$, and our models have a 
sound speed chosen from one of two values, so unsurprisingly, CMOs in
galaxies with the same sound speed have a similar value of $M_0$. However, we
find that $M_{\rm acc} \gg M_0$; that is, subsequent accretion plays a much more
important role in determining the final mass of the CMO than initial
collapse, as shown in Figure~\ref{fig_macc_mgal} ({\em top panel}).

Figure~\ref{fig_macc_mgal} shows that $M_{\rm acc}$ and the average
accretion rate of the CMO increase on average with the total mass
of the galaxy $M_{\rm tot}$ ({\em middle} and {\em bottom} panels). We expect
the CMOs to accrete at roughly the \citet{Bondi1952} rate (eq.~\ref{eq_bondi}),
so a more massive or gas-rich galaxy that has higher gas density, which
enables higher accretion rate, ought to have higher values of $\MCMO$.  

Indeed, at fixed CMO mass and sound speed, equation~(\ref{eq_bondi})
shows $\dot M_{\rm BH} \propto \rho$. In a rotating disk in
hydrostatic equilibrium, the characteristic density scales as $\rho
\propto M_{\rm tot}^{2/3}$ \citep{Wood2000}, assuming that the disk
mass and scale radius are fixed fractions of the total galaxy mass
$M_{\rm tot}$ and size $R\propto M_{\rm tot}^{1/3}$.  Hence, we
expect $\dot M_{\rm BH} \propto M_{\rm tot}^{2/3}$, close to the power-law 
index of 0.6 seen in the bottom panel in Figure~\ref{fig_macc_mgal}. 
On the other hand, the gas-depletion time-scale $\tau_d$, which halts both 
star-formation and the growth of the CMO, depends weakly on $\Mgal$ (see
Fig.~\ref{fig_mcmo_t}). This leads us to a qualitative understanding of the
$\mcmo$ correlation seen in Figure~\ref{fig_mcmo_mgal_sp}. Since the CMO mass
is dominated by accretion, we can very roughly approximate it as $M_{CMO}
\simeq \dot M_{\rm BH} \tau_d \propto M_{\rm tot}^{2/3}$, while $M_{\rm
gal,disk}$ is roughly dependent on the stellar mass of the galaxy. As shown in
Figure~\ref{fig_mstar_mgal}, $M_{\rm star}$ increases linearly with $M_{\rm
  tot}$, $M_{\rm star} \propto M_{\rm tot}$.  This
generally supports the observed positive correlation of $\mcmo$. Although this
argument does not fully explain the observed linear relationship, it does
ignore several factors that likely depend at least weakly 
on $M_{\rm tot}$, including the dark matter contribution
to the galactic disk.

The final mass of the CMO correlates tightly with the {\em global} star
formation efficiency of the galaxy, with a normalization dependent on the
effective sound speed, as shown in Figure~\ref{fig_mcmo_e}. This further
demonstrates a close link between the growth of the CMO and the build-up of
the host galaxy through star formation. The dependence of the CMO mass 
on the effective sound speed can be understood from Bondi-Hoyle accretion.
The initial mass of the CMO is close to the Jeans mass of the collapsing gas,
$M_{\rm {CMO}} \propto M_{\rm J} \propto \rho^{-1/2}c_{\rm s}^3$, so
$\dot{M}_{\rm {BH}} \propto c_{\rm s}^3$. Because of this strong dependence on
the sound speed, our choice of two discrete values of $c_s$ at extremes of the
expected range produces the discrete behavior seen in Figure~\ref{fig_mcmo_e}.
In reality, we would expect a scatter between these extremes.

\section{Central Massive Objects in Galaxy Mergers}

\begin{figure}
\begin{center}
\includegraphics[width=3.5in]{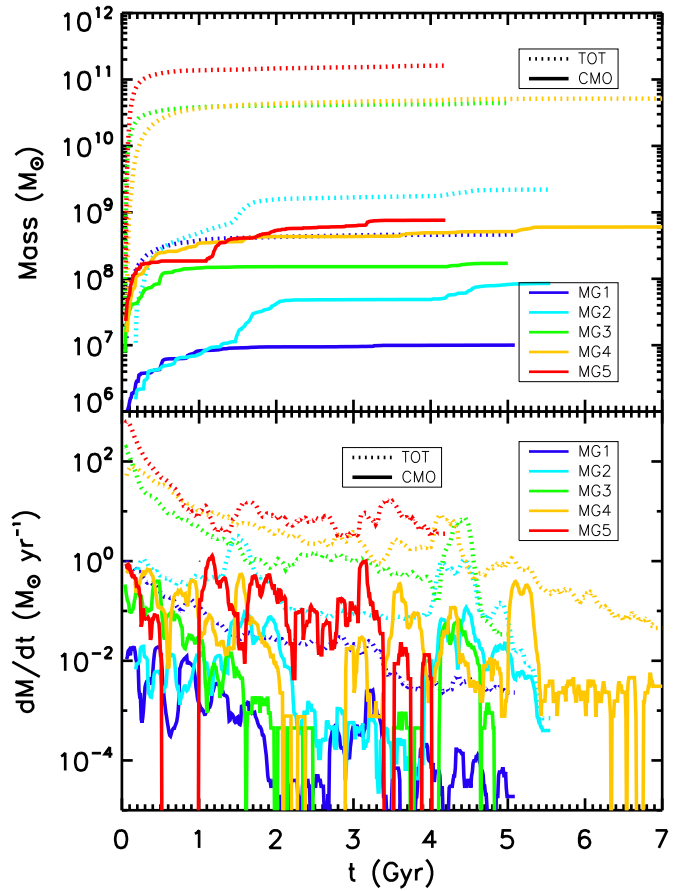}
\vspace{1cm}
\caption{\label{fig_cmo_merger} History of CMO ({\em solid}) and total
  collapsed ({\em dotted}) mass ({\em top}) and accretion rate ({\em bottom})
  in all five galaxy mergers. Total mass collapsed counts mass in all sink
  particles, a quantity related to the total amount of star formation in the
  galaxy by the local star formation efficiency.} 
\end{center}
\end{figure}

\begin{figure}
\begin{center}
\includegraphics[width=3.5in]{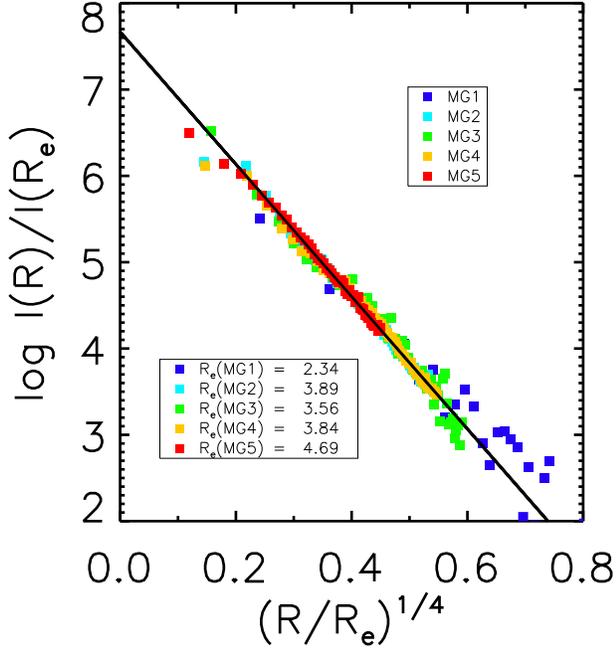}
\vspace{\baselineskip}
\caption{\label{fig_sb_merger} Surface brightness profiles of all
  five merger remnants at the end of the simulations indicated
  in Figure~\ref{fig_cmo_merger}. All five models fit well to a 
\citep{Sersic1963, Sersic1968} profile with $m=4$ ({\em black line}), as
explained in the text. The legend gives the effective radius $\re$ of each
  model. } 
\end{center}
\end{figure}

\begin{figure}
\begin{center}
\includegraphics[width=3.5in]{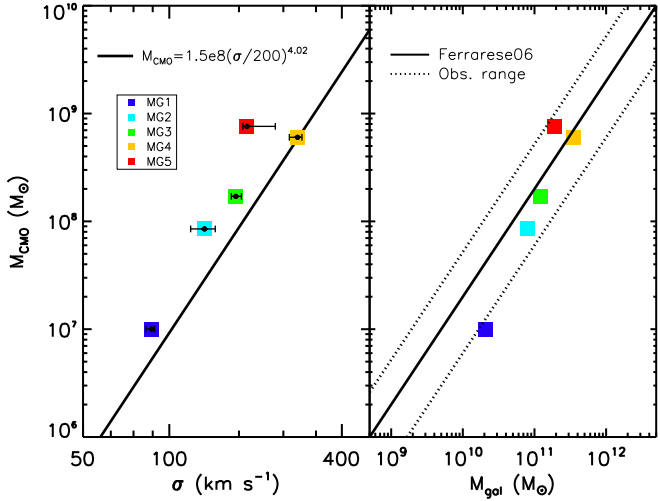}
\vspace{\baselineskip}
\caption{\label{fig_mbh_merger} The relations between CMO mass and
  host mass and velocity dispersion at the conclusion of our galaxy
  merger models. The value of $\MCMO$ is the final mass of the CMO after
  completion of mergers, the $\sigma$ is taken at the mean values,
  and the error bar indicates the range over $10^4$ viewing angles.
  The calculation of $\Mgal$ is described in the text. The black curve
  in the left panel indicates the best fit to observations by
  \cite{Tremaine2002}, while those in the right panel are from observations
  by \cite{Ferrarese2006}.}
\end{center}
\end{figure}

Previous simulations have demonstrated that major mergers of spiral galaxies
generally trigger starbursts and produce elliptical galaxies (e.g.,
\citealt{Toomre1972, Hernquist1989, Barnes1992, 
Hernquist1992, Mihos1996}; \citealt*{Dubinski1999, Springel2000, Barnes2002,
  Naab2003}; \citealt{Li2004}). Recently, simulations of galaxy mergers that
include SMBHs have been carried out to investigate the SMBH-spheroid
relations. Although some dissipationless simulations showed that the $\msigma$
relation could be maintained for mergers with modest energy and angular momentum
  \citep{Nipoti2003, Boylan2005}, it was pointed out that gas dissipation and
  star formation are essential to this correlation \citep{Kazantzidis2005}. In
  particular, \citet{DiMatteo2005} suggest that thermal feedback from an
SMBH suppresses both black hole accretion and star formation,
giving the $\msigma$ correlation in ellipticals.
Together, these studies suggest that the SMBH-spheroid
correlations are closely connected to star formation, and that it may play a
dominant role in determining the masses of SMBH and spheroids. 

In previous simulations, SMBHs are initialized in the progenitors following
the $\msigma$ correlation. In our merger simulations, CMOs form dynamically as
the galaxies evolve,  in the same manner as in our models of isolated spirals.
The growth histories of the CMOs in our merger models are shown in
Figure~\ref{fig_cmo_merger}. Because only the most massive CMO is considered
in the merger remnants, the CMO in galaxy mergers is identified as the most
massive one. Similar to the isolated disk galaxies, the mass curves of the CMO
increase rapidly in the initial phase. However, the detailed accretion history
is modulated by the interaction between the progenitors. Close encounters
trigger episodes of very high accretion rate, both on local and global scale,
as shown in the bottom panel of Figure~\ref{fig_cmo_merger}. The mass
curves saturate after the completion of the mergers.

In order to investigate the $\mcmosigma$ correlation between the CMOs
and the merger remnant velocity dispersions, we made simulated
observations, following \citet{Gebhardt2000}.  We measured the surface
brightness-weighted, line-of-sight stellar velocity dispersion
$\sigma$ of the spheroid within an aperture of size $\re/8$, where
$\re$ is the effective radius. As mentioned in \cite*{Boylan2005} and
\cite{Robertson2006B}, a standard technique for obtaining 
$\re$ is to fit the projected surface brightness profile, $I(R) =
L(R)/(4\pi R^2)$, to a  \citep{Sersic1963, Sersic1968} profile and
derive the half-light radius:
\begin{equation}
    I(R) = I(\re) \exp\{-b(m)[(R/\re)^{1/m}-1]\}. 
\label{sers}
\end{equation}
where $b(m) \approx 2m-1/3+4/(405m)$, as given by
\citet{Ciotti1999}. 

Bulges and early-type galaxies have surface brightness profiles $I(R)$
that are usually well-approximated by the de Vaucouleurs law
\citep{deVaucouleurs1948} 
\begin{equation}
\log I(R)/I(\re) \propto -(R/\re)^{1/m}
\end{equation}
with S{\'e}rsic index $m=4$.  Figure~\ref{fig_sb_merger} shows the surface
brightness profiles $I(R)$ of all three of our merger remnants, assuming a fixed
mass-to-light ratio of three. These profiles appear to follow the
$m=4$ de Vaucouleurs law \citep{deVaucouleurs1948} closely, as also
found by \cite{Hernquist1992} who studied mergers of bulgeless spiral
progenitors. We note that \cite{Balcells2003} and \cite{Graham2005} proposed a  
smaller index, $m \sim 1.7 - 3.5$ for bulges, and it was suggested that only
  ellipticals with $M_{\rm B} \le -21$ mag have $m \simeq 4$ profiles (e.g, 
  \citealt{Graham2003}). However, the derived value of $m$ from fitting
  observations appears to depend on the 
data set and the observed wavelengths. As noted in \cite{Balcells2003}, data
from HST seems to give a much smaller $m$ (e.g., $m \sim 1.7$) than that from
{\em Spitzer} (e.g., $m \sim 4-6$) for the same sample, and low S{\'e}rsic
indices are not expected from galaxy mergers.  \cite{Graham2006} reported a
correlation between the masses of CMOs and the S{\'e}rsic index of the host
spheroids in the sample compiled by \cite{Ferrarese2006}. More merger
simulations with different initial conditions, mass ratios and orbital
parameters are necessary to explore this relation.

The effective radius $\re$ is then derived from the half-light radius of each
spheroid, as given in the legend of Figure~\ref{fig_sb_merger}. 
The mass-weighted line-of-sight stellar velocity dispersion $\sigma$ within
$\re/8$ is then calculated for each remnant galaxy for 10$^4$ random viewing
angles. Figure~\ref{fig_mbh_merger} shows the resulting correlation of
$\mcmosigma$ for the merger remnants. It appears to agree very well with the
observational correlation, $\MBH = 1.5\times 10^8\, (\sigma/200\, \rm km\,
s^{-1})^{4.02}\, \Msun$ (\citealt{Ferrarese2000, Tremaine2002}).

Figure~\ref{fig_mbh_merger} also shows the correlation $\mcmo$ between
the mass of CMO and $\Mgal$ of the host for the galaxy mergers. Here
$\Mgal$ is calculated using the virial mass, following \cite{Cappellari2006},
$\Mgal = \alpha \, \re \, \sigma^2\, /G$, where $G$ is the
gravitational constant and $\alpha = 5$.  This is the best-fitting
virial relation based on the observables $\re$ and $\sigma$. The
simulation agrees very well with the observed correlation $\mcmogal$
(\citealt{Ferrarese2006}), within the $1 \, \sigma$ observational
uncertainty, as indicated by the dotted lines in this figure.

\section{DISCUSSION AND SUMMARY}

Our simulations of gravitational collapse of gas in galaxies enable us
to model star formation in galaxies and follow the growth and
evolution of CMOs. Our approach differs from previous studies such as
\cite{Kazantzidis2005} and \cite{DiMatteo2005} in the following ways.

\begin{enumerate}
\item Our resolution is sufficient to fully resolve
gravitational collapse.
\item Absorbing sink particles are used to directly follow and measure
the mass of gravitationally collapsing gas.
\item In our model, the black holes, or CMOs, are not set up {\em a
priori} according to the $\msigma$ correlation as in previous
approaches. Instead they form dynamically from gravitational collapse
of gas, the same as star clusters, with both being represented by sink
particles. In these simulations, the CMO is just the most massive sink
particle, which always is found in the galaxy center at the end of our
simulations. 
\item Our model does not explicitly include feedback from either star
  formation or black hole formation. However, feedback from star
  formation is implicitly represented by our isothermal equation of
  state with relatively high effective sound speed of 6--15~km~s$^{-1}$.
\end{enumerate}

We emphasize that our models do not include explicit feedback,
magnetic fields, or gas recycling. However, we believe each will have
minor effects on the final mass of the CMO. The assumption of an
isothermal equation of state for the gas implies substantial feedback
to maintain the effective temperature of the gas against radiative
cooling and turbulent dissipation. Real interstellar gas has a wide
range of temperatures but the root-mean-square (rms) velocity dispersion
generally falls within the range 6--15 km s$^{-1}$ (e.g.,
\citealt{Elmegreen2004, Dib2006}).   

At least three effects may conspire to maintain the velocity dispersion in the
observed narrow range, as reviewed in \cite{Li2005A}. First, radiative
cooling drops precipitously in gas with sound speed $\lesssim$ 10 km s$^{-1}$
as it becomes increasingly difficult to excite the Lyman $\alpha$ line of
hydrogen. Second, direct feedback from the starburst may play only a minor
role in quenching subsequent star formation \citep[e.g.][]{kravtsov03, monaco04},
  perhaps because most energy is deposited not in the disk but  
above it as superbubbles blow out \citep[e.g.][]{fujita03,avillez04}. The
energy input $\dot{E}$ from supernovae at the observed Galactic rate drives a
flow with rms velocity dispersion of $\sim$ 9.5 km s$^{-1}$
(\citealt{avillez00}), and the rms velocity dispersion depends only on 
$\dot{E}^{1/3}$ (\citealt{maclow99, mk04}), so a wide range of star formation
rates leads to a narrow range of velocity dispersions.  Third,
magnetorotational instabilities may maintain a modest velocity dispersion
even in the absence of any other feedback (\citealt*{sellwood99, dzi04}).

Radiative feedback from the central SMBH in massive galaxies will only Compton
heat the gas much above $10^4$~K if it the gas is sufficiently ionized.
\citet{Sazonov2005} shows that this will occur if the ionization
parameter 

\begin{equation}
\xi = \frac{L}{nr^2} = 1.4 \times 10^8 \frac{\MBH}{10^8 {\rm
    M}_{\odot}} \frac{1 \mbox{ cm}^{-3}}{n} \left(\frac{1 \mbox{
    pc}}{r} \right)^2\,,
\end{equation}
exceeds $\xi_{\rm crit} \simeq 10^3$, with the Compton temperature of
$\sim 10^7$~K being reached only for $\xi \simeq 10^5$. In this
equation, $\MBH$ is the SMBH mass, $L$ is the bolometric luminosity,
which is assumed to be a constant fraction one-tenth of the Eddington
luminosity, $n$ the number density, and $r$ the distance from the
SMBH.  If we examine the conditions at the Bondi-Hoyle accretion
radius for the SMBH,

\begin{equation}
R_{\rm BH} = \frac{G \MBH}{c_s^2} \simeq (4.5 \mbox{ kpc}) \frac{\MBH}{10^8 {\rm
    M}_{\odot}} \left(\frac{c_s}{10\mbox{ km s}^{-1}}\right)^{-2}
\end{equation}
we find that gas with a density of $n > (9 \times 10^{-3} \mbox{
cm}^{-3}) (\MBH / 10^8 {\rm M}_{\odot})$ will remain at $10^4$~K (see
also \citealt{Sazonov2005}). This may contribute to maintaining the
sound speed at roughly 10~km~s$^{-1}$ in dense gas further in that
might otherwise cool.  On the other hand, the heating in the immediate
neighborhood of the SMBH may still lead to unsteady accretion as
modeled by \citet{Ciotti2001}.

We should also emphasize that gas recycling
from dying stars is not included in our simulations. As computed
explicitly, for example, in \cite{Haiman2004B}, stars can return over
40\% of their mass in gas. However, since the recycled gas might be
heated up or blown out by the feedback, it may not be accreted by the
CMOs. Our models with different effective sound speeds satisfy
similar correlations between $\MCMO$ and host galaxy mass (though not
the same relation between CMO mass and star formation
efficiency). Furthermore, as discussed in \S~2.2, the mass loss due to
galactic winds is only a small fraction of the initial gas mass,
suggesting that the effect of feedback on the mass of the CMO may be
minor.

The simple model of sink particles that grow with approximately
spherical Bondi-Hoyle accretion reproduces the observed $\mcmosigma$ and
$\mcmo$ correlations reasonably. Both the accreted mass of CMO and the mass
turned into stars scale with the total mass of a galaxy, offering a plausible
explanation for the CMO-host relation. We should note that \cite{Escala2006}
also reproduces the $\msigma$ correlation by modeling accretion of BHs from
a Shakura-Sunyaev thin disk \citep{Shakura1973}. This suggests that the
accretion rates through these two different mechanisms may be similar, and
that the final mass of the black hole or the CMO may be largely determined by
the depletion of accretable gas. The fact that various previous approaches,
including feedback models (e.g. \citealt{DiMatteo2005, Sazonov2005}), star
formation models (e.g., \citealt{Kazantzidis2005}), and accretion models
(e.g., \citealt{Escala2006}), have more or less succeeded in reproducing the
SMBH-host correlation suggests that gas depletion in the central region is
crucial to this correlation. Our results support the idea that star formation
plays the major role in gas consumption and determination of the final masses
of the CMO and the stellar component.  

The close correlation that we find between the CMO mass and global
star formation efficiency in the host galaxy suggests that the
CMO-host link may be universal, in that it is the result of the
co-eval growth and evolution of the CMO and the host galaxy, and
therefore it does not strongly depend on the morphology, type, or mass
of the galaxy. A systematic search for CMOs in the nuclei of bulgeless
disk galaxies would offer a test of this conclusion.

In summary, our simple model of accretion and star formation reproduces
quantitatively the observed $\msigma$ correlation for galaxy mergers, and
suggests a universal $\mcmo$ relation over a wide range of galaxy mass and
morphological types. We find that the CMO builds up its mass through 
accretion, and that there is a direct correlation between the global
star formation efficiency of a galaxy and its CMO mass. Our results suggest 
that star formation may play an important role in producing the fundamental
mass correlation between the central massive objects and their host galaxies. 

\acknowledgments{We thank V. Springel for making both GADGET and
his galaxy initial condition generator available, as well as K.
Gebhardt, A. Graham, L. Hernquist, A. Loeb, J. Ostriker, and E.
Quataert for useful discussions and comments on the manuscript.  YL thanks
L. Hernquist for his 
encouragement, and is supported  
by an Institute for Theory and Computation Fellowship.
ZH acknowledges partial support by NASA through grants NNG04GI88G and
NNG05GF14G, by the NSF through grants AST-0307291 and AST-0307200, and by the
Hungarian Ministry of Education through a Gy\"orgy B\'ek\'esy Fellowship. 
M-MML acknowledges partial support by NASA under grant NAG5-13028,
and by the NSF under grants AST-9985392 and AST-0307854.
Computations were performed at the Pittsburgh Supercomputer
Center supported by the NSF, on the Parallel Computing Facility of the
AMNH, and on an Ultrasparc III cluster generously donated by Sun
Microsystems.}


\end{document}